\documentclass[aps,prd,a4paper,longbibliography,reprint,superscriptaddress,floatfix,nofootinbib,amsmath,amssymb,preprintnumbers,showpacs,showkeys,fleqn]{revtex4-2}
\pdfoutput=1   
\usepackage{amsmath,amsfonts,amssymb,amsbsy}
\usepackage{mathrsfs,latexsym}
\usepackage{graphicx}
\usepackage{xcolor}
\usepackage[bookmarks,
bookmarksopen = true,
bookmarksnumbered = true,
linktocpage,
colorlinks = true,
linkcolor = blue,
urlcolor  = blue,
citecolor = blue,
anchorcolor = green,
hyperindex = true,
hyperfigures]{hyperref}
\usepackage{array}
\usepackage{booktabs}
\usepackage{multirow}
\usepackage{subfig}
\usepackage{placeins}
\usepackage{blindtext}
\usepackage{appendix}
\usepackage{bm}
\usepackage{times}
\usepackage{float}
\usepackage{wrapfig}
\usepackage[utf8]{inputenc} 
\usepackage[T1]{fontenc} 
\usepackage{amsmath,amssymb}

\newcommand{\eq}[1]{Eq.~(\ref{#1})}

\newcommand{\sect}[1]{section~\ref{#1}}

\newcommand{\fig}[1]{figure~\ref{#1}}

\newcommand{\tab}[1]{table~\ref{#1}}

\newcommand{\ben}{\begin{enumerate}}
\newcommand{\een}{\end{enumerate}}
\newcommand{\bit}{\begin{itemize}}
\newcommand{\eit}{\end{itemize}}
\newcommand{\nn}{\nonumber \\}
\newcommand{\non}{\nonumber}

\begin{document}

\title{Constructing static quark-anti-quark
creation operators from Laplacian eigenmodes\\[2ex]
}


\author{Roman~H\"ollwieser} 
\author{Francesco Knechtli}
\author{Tomasz Korzec}
\affiliation{Department of Physics, University of Wuppertal, Gau{\ss}strasse 20, 42119 Germany}
\author{Michael Peardon}
\affiliation{School of Mathematics, Trinity College Dublin, Ireland}
\author{Juan Andrés Urrea-Niño}
\affiliation{Department of Physics, University of Wuppertal, Gau{\ss}strasse 20, 42119 Germany}

\date[]{ November 2022}

\begin{abstract}
We investigate static quark anti-quark operators based on trial states formed 
from eigenvectors of the covariant three-dimensional lattice Laplace operator. 
We test the method by computing the static quark-anti-quark potential and comparing results to standard Wilson loop measurements. The new method is efficient not only for on-axis, but also for many off-axis quark-anti-quark separations when a fine spatial resolution is required. We further improve the ground-state overlap by using multiple eigenvector pairs, weighted with Gaussian profile functions of the eigenvalues, providing a variational basis. The method presented here can be applied to  potential functions for all possible excitations of a gluonic string with fixed ends, hybrid or tetra-quark potentials, as well as static-light systems and allows visualization of the spatial distribution of the Laplace trial states. 
  \end{abstract}
  

\preprint{WUB/22-05}


\maketitle


\newpage

\section{Introduction}

The potential of a static quark-anti-quark pair $V_{0}(r)$ has always played an important role in Quantum Chromodynamics (QCD). It can be computed via Wilson loops~\cite{Wilson:1974sk} and established an understanding of confinement and its interplay with asymptotic freedom, a central problem of particle physics, via the formation of a flux tube between quark-anti-quark static charges~\cite{DiGiacomo:1989yp, DiGiacomo:1990hc, Singh:1993jj, Bali:1994de, Bali:2000gf, Luscher:2002qv,Greensite:2005yu,Andreev:2020pqy}. Confinement manifests itself in the linear rise of $V_{0}(r)$ at large $r$; the corresponding slope is known as the string tension. 
The static potential can be used in the Born-Oppenheimer approximation~\cite{Born:1927opp} to compute the spectrum of quarkonium~\cite{Campbell:1987nv,Perantonis:1990dy,Bali:2000vr,Braaten:2014qka,Capitani:2018rox,Bicudo:2020qhp}.
It is also an important observable in setting the scale in lattice QCD. In quenched calculations, the scale has been set using the string tension, but in full QCD the string breaks at the pair-production threshold, making a precise definition difficult. The static energy allows determination of the strong coupling, $\alpha_s$, or, equivalently, $\Lambda_{\scriptstyle \overline{\textrm{MS}}}$; see~Refs.~\cite{DallaBrida:2020pag,dEnterria:2022hzv} for recent reviews. Instead of the static energy, one can also use the force $F(r) \equiv dV_{0}(r)/dr$, which is free of the self-energy linear divergence. The dimensionless product $r^{2}F(r)$ can be used to set the scale~\cite{Sommer:1993ce} at distances where statistical and systematic uncertainties are under good control, {\it e.g.}, $r_{0}$ or $r_{1}$, 
defined by $r_{i}^{2} F(r_{i})= c_{i}$, with $c_{0}=1.65$~\cite{Sommer:1993ce}, $c_{1}=1$~\cite{Bernard:2000gd}.

In this paper, we investigate a method for computing the static quark-anti-quark potential in lattice QCD not based on Wilson loops, but where trial states are formed from components of eigenvectors of the covariant lattice Laplace operator~\cite{Neitzel:2016lmu}. In this construction, the spatial Wilson lines in the Wilson loop are replaced by outer products of Laplacian eigenvectors. This idea was proposed in the context of adjoint string breaking \cite{deForcrand:1999kr} and of Polyakov loops and the static potential at finite temperature \cite{Philipsen:2002az, Jahn:2004qr}. The main advantage is we can not only form straight lines (on-axis), but also off-axis paths very easily. These correspond to very complicated stair-like constructions of spatial link variables. It is important to compute the static potential for many off-axis separations whenever a fine resolution is required, {\it e.g.}, for a detailed investigation of string breaking \cite{Bali:2005fu, Bulava:2019iut} or to determine the scale $\Lambda_{\scriptstyle \overline{\textrm{MS}}}$ via matching the perturbative and the lattice QCD static potential~\cite{Brambilla:2010pp, Jansen:2011vv, Bazavov:2012ka, Bazavov:2014soa}. It is even mandatory to compute all possible on- and off-axis separations to determine the static potential in momentum space representation~\cite{Karbstein:2014bsa}. 

The implementation of~\cite{Neitzel:2016lmu} which uses only the eigenvector corresponding to the lowest eigenvalue can be significantly improved by summing over several eigenvectors, weighted by Gaussian profile functions of their corresponding eigenvalues. A similar method was successfully applied to hadronic correlation functions in~\cite{Knechtli:2022bji} where an optimal smearing profile was introduced in the distillation framework~\cite{Peardon:2009gh}, which can be equivalently expressed as an optimal creation operator for a meson. In the case of the static potential we get an improvement for the static energies, which reach their plateau values at earlier temporal distances, to be quantified below. The improved implementation can also be adapted to measure multi-quark potentials, hybrid static potentials of exotic mesons, where the gluonic string excitations can be realized by applying covariant derivatives to the Laplacian eigenvectors, as well as static-light potentials with insertions of light quark propagators. Further, we present a simple way to illustrate the flux tube between a static quark and antiquark pair using a Laplacian eigenvector pair as a 'test charge' scanning the chromo-electromagnetic field.

The article is organized as follows: First, we reintroduce the notation of {\it Laplace trial states} in~\sect{sec:lts}. Next we reformulate the standard Wilson loop in terms of {\it Laplace trial state correlators} and discuss their improvement via Gaussian profile functions in ~\sect{sec:imp}, allowing us to formulate a generalized eigenvalue problem (GEVP) for the {\it Laplace trial state correlation basis matrix} of the static potential, resulting in optimal profile functions for ground and excited states. We test the new improved method on a dynamical fermion ensemble in~\sect{sec:res}, presenting results for effective  energies, static potentials as well as excited states. In~\sect{sec:flux} we look at the spatial distribution of the optimal Laplace trial states which probe the ground and excited static potentials of a quark-anti-quark pair. We draw our conclusions and give a short outlook in~\sect{sec:co}. 

\bigskip

\section{Laplace trial states}\label{sec:lts}

Let ${\bar Q}^a(\vec x)$ denote a static color source with $a=1,2,3$ at spatial position $\vec x$. Wilson loops arise from correlations in time of trial states $\bar{Q}(\vec x) U_s(\vec x, \vec y) Q(\vec y)$ for a static color anti-color source pair located at spatial positions $\vec x$ and $\vec y$ respectively\footnote{We omit the time coordinate in this section since trial states exist on single time-slices only.}. Note the same Wilson loops are obtained when the static color sources are replaced by static quarks since the heavy quark spins decouple in the static limit and the trace over spin yields a constant, see \cite{Donnellan:2010mx}. The spatial Wilson line $U_s(\vec x, \vec y)=\exp(i\int_{\vec x}^{\vec y}A_\mu dx^\mu)=\prod U_\mu$ is a path-ordered product of link variables from $\vec x$ to $\vec y$. We want to replace the spatial part of trial states in each time-slice with an alternative operator which respects the gauge transformation behavior of the spatial Wilson line, given by 
\begin{eqnarray}
U_s'(\vec x, \vec y) = G(\vec x)U_s(\vec x, \vec y)G^\dagger(\vec y),
\end{eqnarray}
to ensure gauge invariance of the trial state. 

\begin{figure*}[ht]
\centering
\includegraphics[width=0.66\textwidth]{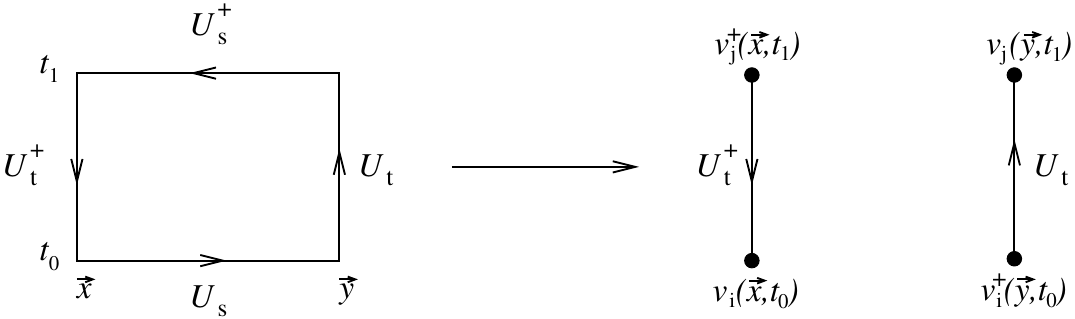}
\caption{The spatial Wilson lines $U_s(\vec x,\vec y,t)$ of the classical Wilson loop $W(R,T)$ of size $(R=|\vec y-\vec x|)\times(T=|t_1-t_0|)$ (left) can be replaced by Laplacian eigenvector pairs $v_i(\vec x,t)v_i^\dagger(\vec y,t)$ (right, eigenvector pairs to be read in anti-clockwise direction), to form Laplace trial state correlators via the two static perambulators $\bar\tau_{ij}(\vec x,t_0,t_1)$ and $\tau_{ij}(\vec y,t_0,t_1)$.}
  \label{fig:wlapl}
\end{figure*}

The three-dimensional gauge-covariant lattice Laplace operator $\Delta$, acting on a field $\psi(\vec x)$ on a single time-slice of the four-dimensional 
lattice gives 
\begin{eqnarray}
\Delta \psi(\vec x)&=&\frac{1}{a^2}\sum_{k=1}^3[U_k^\dagger(\vec x-a\hat k)\psi(\vec x-a\hat k)\\
&&\qquad\qquad-2\psi(\vec x)+U_k(\vec x)\psi(\vec x+a\hat k)]\non
\end{eqnarray}
and has the required transformation behavior $\Delta'(\vec x,\vec y)=G(\vec x)\Delta(\vec x,\vec y)G^\dagger(\vec y)$. Eigenvalues $\lambda$ of $\Delta$ are gauge invariant, while eigenvectors $v'(\vec x)=G(\vec x)v(\vec x)$ transform co-variantly~\cite{Bruckmann:2005hy}. 
It follows, that we can write down a combination of eigenvector components for a given eigenvalue $\lambda$, namely $v(\vec x)v^\dagger(\vec y)$, which has the same behavior under gauge transformations as the spatial Wilson line $U_s(\vec x,\vec y)$:
\[v'(\vec x)v'^\dagger(\vec y)=G(\vec x)v(\vec x)v^\dagger(\vec y)G^\dagger(\vec y).\]

At this point, inspired by the distillation operator \cite{Peardon:2009gh}
\begin{eqnarray}
\square^{ab}(\vec z , \vec x) & = & \sum_{i=1}^{N_v}v_i^a(\vec z)v_i^{\dagger\,b}(\vec x) \,,\label{eq:box}
\end{eqnarray}
we introduce the more general operator 
\begin{eqnarray}
\tilde\square^{ab}(\vec z , \vec x) & = & \sum_{i=1}^{N_v}\rho_i v_i^a(\vec z)v_i^{\dagger\,b}(\vec x) \,,\label{eq:mbox}
\end{eqnarray}
by including a quark profile $\rho_i$, which modulates contribution from different eigenmodes. Note $\square$ is a projection matrix, $\square^2=\square$ onto 
$V$, the vector space spanned by $\{v_i\}$, while $\tilde\square$ is no longer idempotent, it still has an image given by the span of $v_i$. Next, we define 
the auxiliary field on each time-slice 
\begin{eqnarray}
\chi^a(\vec z | \vec x) & = &\tilde\square^{ab}(\vec z , \vec x) Q^b(\vec x)\quad(\text{no sum over }\vec x)\nn
&=& \sum_{i=1}^{N_v} \rho_i v_i^a(\vec z)v_i^{\dagger\,b}(\vec x) Q^b(\vec x).
\label{chifield}
\end{eqnarray}
$\chi^a(\vec z | \vec x)$ can be interpreted as an effective smeared color-electromagnetic field over the whole time-slice induced by the static source at $\vec x$.  At first this seems contradictory to a 'static' color source, but it follows the notation of distillation. We stress the role of the 'smearing parameter' $N_v$, the number of eigenvectors to be summed over in \eq{eq:box}, behaves opposite to intuition. $N_v=1$ corresponds to the maximal smearing and in the limit where all eigenvectors are included $N_v\rightarrow3N_s^3$ with $N_s^3$ the spatial lattice volume of a time slice, the smearing operator becomes the identity. This we have to keep in mind when constructing gauge invariant trial states for a color anti-color source pair located at spatial positions $\vec x$ and $\vec y$, respectively, via
\begin{eqnarray}
\Phi(\vec x, \vec y) & = & \sum_{\vec z} \bar{\chi}(\vec z | \vec x) \chi(\vec z | \vec y)\nn   
& = &\bar{Q}(\vec x)\sum_{i,j=1}^{N_v}\rho_i\rho_jv_i(\vec x)\nn
&&\quad\quad\;\;\underbrace{\sum_{\vec z}v_i^\dagger(\vec z)v_j(\vec z)}_{=\delta_{ij}}v_j^\dagger(\vec y) Q(\vec y)\nn
& = &\bar{Q}(\vec x)\sum_{i=1}^{N_v}\rho_i^2v_i(\vec x)v_i^\dagger(\vec y) Q(\vec y) \,,
\label{eq:lts}
\end{eqnarray}
where we used the orthonormality of the Laplacian eigenvectors, which ensures that the standard distillation operator is idempotent, {\it i.e.}, $\square^2 = \square$. We denote \eq{eq:lts} as a {\it Laplace trial state}, the positions $\vec x$ and $\vec y$ label the sector of the Hilbert space in which the static energies will be determined.

Notice the sum over eigenvectors in \eq{eq:lts} must be truncated at finite $N_v$ or a non-trivial profile $\rho_i$ must be applied to avoid the collapse of the {\it Laplace trial state}, or
the annihilation of the quark-anti-quark pair. For example $\rho_i=\delta_{ik}$ corresponds to the choice of a single eigenvector $v_k$. A simple truncation of the sum at some finite $N_v=k$ could be formulated via $\rho_i=\Theta(k-i)$ and we can of course introduce multiple profile functions to define an operator basis $\Psi^{(k)}$ via different profiles $\rho_i^{(k)}$ For example, $\Psi^{(k)}$ with $\rho_i^{(k)}=e^{-\lambda_i^2/4\sigma_k^2}$ corresponds to a sum over eigenvectors weighted with Gaussian profiles in eigenvalue space with different Gaussian widths $\sigma_k$, which turned out to be very efficient for meson operators in \cite{Knechtli:2022bji}. In the following section we will reformulate the usual Wilson loops in terms of  {\it Laplace trial state correlators} and follow the same strategy as in \cite{Knechtli:2022bji} by introducing a set of Gaussian profile functions into the the correlators and solving a generalized eigenvalue problem (GEVP) for the {\it Laplace trial state correlation matrix} to extract optimal trial state profiles $\tilde\rho_i^{(n)}$ for ground and excited states of the static potential $V_n(R)$, ($n=0,1,2\ldots$). We also tried other profile functions, {\it e.g.}, $\delta$- or $\Theta$-functions to construct an $N_v\times N_v$ transfer matrix with individual eigenmode pair contributions or summing up different numbers of eigenmodes $N_v$ to construct a GEVP basis matrix like the ordinary construction using Wilson loops with different spatial smearing levels. Different profiles yield the same results, yet the Gaussian basis seems the most natural (vs. $\delta$- or step-functions) and numerically stable choice.

\section{The static quark-anti-quark potential from Laplace trial state correlators}\label{sec:imp}

\begin{figure}[h]
\centering
\includegraphics[width=0.495\textwidth]{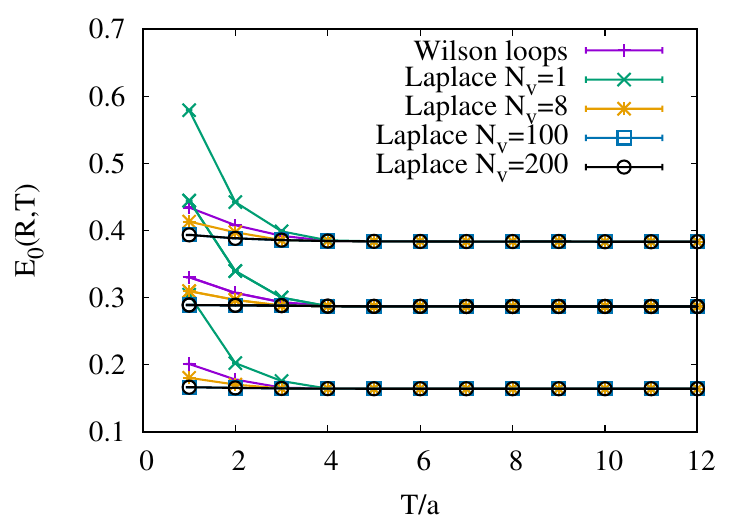}
\caption{The effective energies for $R/a=2-4$ from Wilson loops and Laplace trial state correlators with increasing num- bers of eigenvectors $N_v$. The ground state overlap drastically improves by using more eigenvectors, we see earlier plateaus for larger $N_v$, also quantified in \tab{tab:ovl}. The lines connecting the measured points just help to guide the eye.}
  \label{fig:vmeffEm1}
\end{figure}

The standard Wilson loop $W(R,T)$ of size $(R=|\vec r|=|\vec y-\vec x|)\times(T=|t_1-t_0|)$ can be rewritten using Laplace trial state correlators by replacing the spatial Wilson lines $U_s(\vec x,\vec y,t)$ with Laplacian eigenvector pairs $v_i(\vec x,t)v_i^\dagger(\vec y,t)$, as depicted in \fig{fig:wlapl}. 
The temporal Wilson line $U_t(\vec y,t_0,t_1)$, representing static time-like propagation for a color source at space point $\vec y$ from time $t_0$ to $t_1$ is sandwiched between eigenvectors at the corresponding start- and end-times $v_i^\dagger(\vec y,t_0)$ and $v_j(\vec y,t_1)$. Distinct eigenvector indices appear at the source and sink times, so this can be interpreted as the static 
perambulator
\begin{eqnarray}
\tau_{ij}(\vec y,t_0,t_1)=v_i^\dagger(\vec y,t_0)U_t(\vec y,t_0,t_1)v_j(\vec y,t_1),\label{eq:tau}
\end{eqnarray}
at $\vec y$ of time extent $T=|t_1-t_0|$. Its expectation value $\langle \tau_{ij}(\vec y,t_0,t_1)\rangle$ vanishes of course. When combined with another static perambulator $\tau_{ji}(\vec x,t_1,t_0)$ at $\vec x$, it gives the Laplace trial state correlator 
\begin{eqnarray}
\hspace{-8mm}L(R,T)=\bigg\langle\sum_{i,j}^{N_v}\rho_i^2(t_0)\rho_j^2(t_1)\tau_{ij}(\vec y,t_0,t_1)\tau_{ji}(\vec x,t_1,t_0)\bigg\rangle\label{eq:Lkl}
\end{eqnarray}
for $R=|\vec y-\vec x|$ (in our measurements we average over all $\vec r$ of the same $R$). 
To test the method, the correlation function of \eq{eq:Lkl} is computed on a $N_t\times N_s^3$ lattice ensemble with $N_t=48, N_s=24$ and compared with 
standard Wilson loops. The Wilson loops are determined on 4646 gauge 
configurations while the Laplace trial-state correlators are computed on every 
fourth configuration only to give 1160 measurements. 
We extract the static potential via $aV_0(R)=\lim_{T\rightarrow\infty}\log[L(R,T)/L(R,T+a)]$. First, we analyze the effect of increasing the number of eigenmodes $N_v$ for trivial quark profiles. In \fig{fig:vmeffEm1} we plot the effective energies for the static quark-anti-quark pair for $R/a=2, 3$ and 4, and clearly see an increasing number $N_v$ of Laplacian eigenvector pairs improves the overlap with the ground state drastically. Already $N_v=8$ eigenvector pairs reach the plateau values faster than the original Wilson loops. The improvement seems to saturate at about $N_v\approx100$, we do not see a difference between $N_v=100$ and $N_v=200$. The ground state overlaps can also be quantified by taking the $t$-average over the mass-plateau region of the fractional overlap
\begin{eqnarray}
A_{\rm eff}=\frac{L(R,t)}{L(R,t_S)}\frac{\cosh \left( \left( \frac{aN_t}{2} - t_S \right) aV_0(R) \right)}{\cosh \left( \left( \frac{aN_t}{2} - t \right) aV_0(R) \right)},\label{eq:ovl}
\end{eqnarray}
using the same $t_S=3a$ for all $R/a$ and corresponding ground state energies $aV_0(R)$ from a $\cosh$-fit, for more details see~\cite{Knechtli:2022bji}. These fractional overlaps are listed in \tab{tab:ovl} and demonstrate that a large number $N_v$ of eigenvector pairs gives better overlaps for small distances $R/a$, but with decreasing importance for large distances, where already $N_v<100$ shows better overlaps. 

\begin{table}[h]
\centering
\begin{tabular}{c|ccccc|c}
$R/a$ & $N_v=1$ & $8$ & $100$ & $200$ & optimal & Wloop \\
\midrule
2 & 0.747(4) & 0.929(2) & 0.988(1) & 0.987(1) & 0.989(1) & 0.978(1) \\
3 & 0.723(4) & 0.878(2) & 0.987(2) & 0.986(1) & 0.988(1) & 0.972(2) \\
4 & 0.726(5) & 0.874(3) & 0.982(2) & 0.984(2) & 0.986(2) & 0.965(3) \\
5 & 0.637(6) & 0.871(4) & 0.983(3) & 0.982(3) & 0.983(3) & 0.956(5) \\
6 & 0.629(6) & 0.869(4) & 0.981(4) & 0.980(3) & 0.981(3) & 0.948(6) \\
7 & 0.619(7) & 0.869(5) & 0.982(4) & 0.979(4) & 0.987(4) & 0.934(7) \\
8 & 0.598(8) & 0.862(6) & 0.971(5) & 0.970(4) & 0.974(4) & 0.953(8) \\
9 & 0.572(8) & 0.857(6) & 0.954(5) & 0.934(4) & 0.963(3) & 0.947(9) \\
10 & 0.540(9) & 0.840(7) & 0.941(6) & 0.931(5) & 0.965(1) & 0.94(1) \\
11 & 0.426(9) & 0.807(7) & 0.934(5) & 0.93(1) & 0.956(9) & 0.93(1) \\
12 & 0.33(7) & 0.79(2) & 0.932(9) & 0.92(1) & 0.95(1) & 0.92(1)
 \end{tabular}
 \caption{Fractional overlaps with the corresponding ground state energy $aV_0(R)$ as defined in \eq{eq:ovl}. An increasing number $N_v$ of Laplacian eigenvector pairs enhances the overlap up to about $N_v\approx100$. The overlaps for Laplace trial states from a GEVP with optimal quark profiles in the 6th column are better than standard Wilson loop results from a GEVP with different HYP smearing levels in column 7.}\label{tab:ovl}
 \end{table}

\begin{figure}[h]
\centering
\includegraphics[width=0.45\textwidth]{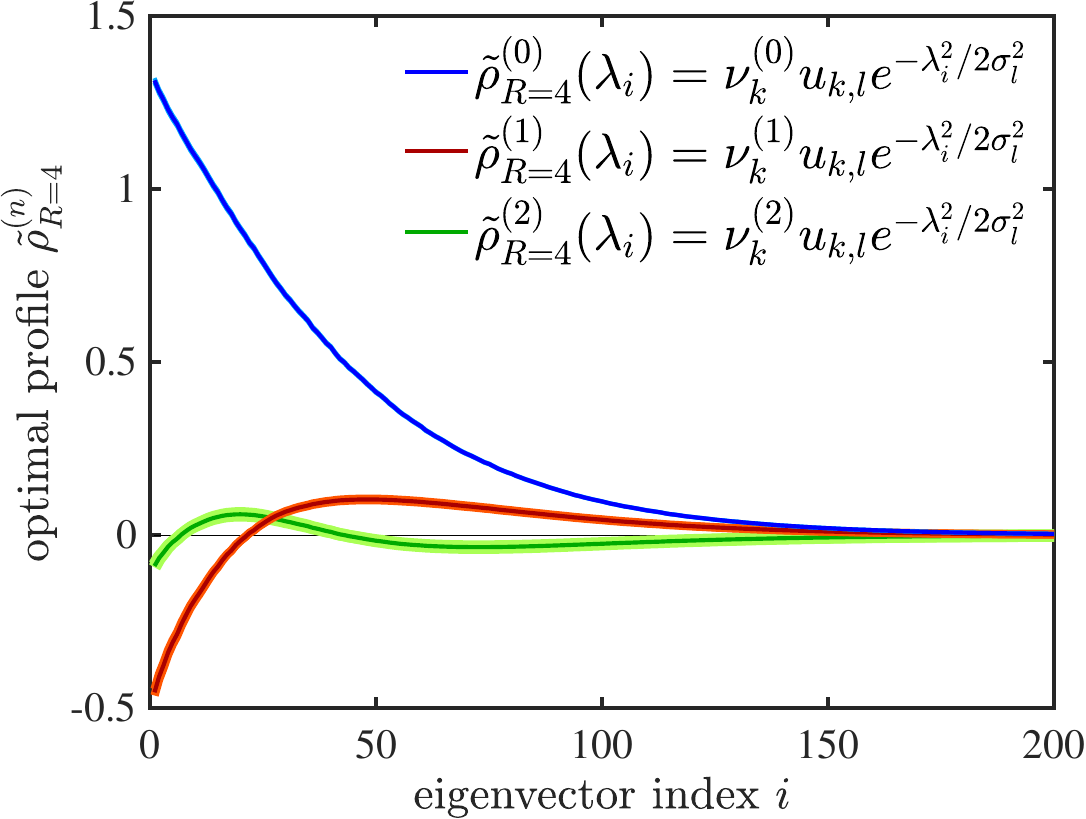}
\caption{The optimal trial state profiles for ground (blue) and excited (red, green) states $\tilde\rho^{(n)}_R(\lambda_i)$, \eq{eq:propti} at $R=4a$.}
  \label{fig:propti}
\end{figure}

Next, instead of trivial quark profiles $\rho_{i,j}$, we use Gaussian quark profile functions $\rho_i^{(k)}=e^{-\lambda_i^2/4\sigma_k^2}$ and $\rho_j^{(l)}=e^{-\lambda_j^2/4\sigma_l^2}$ for the Laplace trial states at $t_0$ and $t_1$ with corresponding eigenvalues $\lambda_{i,j}$ and Gaussian widths $\sigma_{k,l}\in[0.05,0.0894,0.1289,0.1683,0.2078,0.2472,0.2867]$. We define the $7\times7$ Laplace trial state correlation matrix $\mathcal L_{kl}(R,T)$ and solve a generalized eigenvalue problem (GEVP)~\cite{Blossier:2009kd} to identify the optimal trial state profiles $\tilde\rho_R^{(n)}(\lambda)$ for various energy levels $V_n(R)$ ($n=0,1,2,\ldots$). First, we apply the strategy presented in~\cite{Balog:1999ww, Niedermayer:2000yx} and prune $\mathcal L_{kl}$ using the three most significant singular vectors $u_i$
from a singular value decomposition\footnote{$\mathcal L_{kl}=UDV^\dagger$ with $U=V$ (because $\mathcal L$ is Hermitian in our case) being a unitary matrix, whose column vectors $u_i$ form an orthonormal basis, and $D$ being diagonal with non-negative real numbers on the diagonal.} (SVD) at a specific $t_G=4$ via $\tilde{\mathcal L}_{mn}=u_{m}^\dagger\mathcal L_{kl}u_{n}$, which keeps a smaller set of distinct profiles which improves the stability of the GEVP. We perform the latter at the same $t_G$, separately for all spatial distances $R$: 
\begin{eqnarray}
\tilde{\mathcal L}(t)\nu^{(n)}(t,t_G)=\mu^{(n)}(t,t_G)\tilde{\mathcal L}(t_G)\nu^{(n)}(t,t_G).
\end{eqnarray}
From the eigenvalues or so-called principal correlators $\lim_{t\rightarrow\infty}\mu^{(n)}(t,t_G)=e^{-E_n(t-t_G)}$ we get the effective energies for a fixed $t_G$, by performing a $\cosh$-fit in practice, due to periodic boundary conditions. From the generalized eigenvectors $\nu_k^{(n)}$ we can construct the optimal trial state profiles $\tilde\rho^{(n)}_R$ for the energy states provided by the GEVP, which also depend on the quark separation $R$, obviously. First, we use the singular vectors $u_l$ to get the pruned (or most significant) profiles $\bar\rho_R^{(k)}(\lambda_i)=\sum_lu_{k,l}e^{-\lambda_i^2/2\sigma_l^2}$. Then we form the linear combination of pruned profiles using the generalized eigenvectors $\nu_k$ to give the optimal trial state profiles 
\begin{eqnarray}
\tilde\rho^{(n)}_R(\lambda_i)=\sum_k\nu_k^{(n)}\bar\rho_R^{(k)}=\sum_{k,l}\nu_k^{(n)}u_{k,l}e^{-\lambda_i^2/2\sigma_l^2}\,,\label{eq:propti}
\end{eqnarray}
depicted in \fig{fig:propti} for the ground and excited states at $R=4a$. The optimal profiles suggest a number $N_v<100$ of significant/important eigenvectors in the correlator, because each trial state comes with a profile and the combination falls off about twice as fast compared to \fig{fig:propti}. The fractional overlaps with the ground state in \tab{tab:ovl} also favor the Laplace trial states from a GEVP with optimal profiles in the 6th column, which are even better than standard Wilson loop results from a GEVP with different HYP smearing levels (col. 7).


\section{Results from optimal Laplace trial states}\label{sec:res}

We performed all our measurements on $48\times 24^3$ lattices with periodic boundary conditions except for anti-periodic boundary conditions for the fermions in the temporal direction. They were produced with the openQCD package \cite{Luscher:2012av} using the plaquette gauge action and two dynamical non-perturbatively $O(a)$ improved Wilson quarks \cite{Jansen:1998mx} with a mass equal to half of the physical charm quark mass. The bare gauge coupling is $g_0^2=6/5.3$ and the hopping parameter is $\kappa=0.13270$. The scale $r_0/a=4.2866(24)$~\cite{Sommer:1993ce} and the flow scale \cite{Luscher:2010iy} is $t_0/a^2 = 1.8477(3)$. The corresponding lattice spacing is $a=0.0658(10)$fm \cite{Fritzsch:2012wq,Cali:2019enm}. All measurements were performed by our C+MPI based library that facilitates massively parallel QCD calculations. A total of $N_v = 200$ eigenvectors of the 3D covariant Laplacian were calculated on each time-slice of the lattices as described in~\cite{Knechtli:2022bji}. 
A total of 20 3D APE smearing \cite{Albanese1987} steps with $\alpha_{APE} = 0.5$ were applied on each gauge field before the eigenvector calculation so as to smooth the link variables that enter the Laplacian operator. When forming the correlations of the Laplace trial states, we apply one HYP2 smearing step to the temporal links~\cite{Hasenfratz:2001hp, DellaMorte:2003mw, DellaMorte:2005nwx, Grimbach:2008uy, Donnellan:2010mx}. Standard Wilson loops were measured using the wloop package~\cite{wloop}, also applying one HYP2 step to all gauge links, and 4 levels (0 10 20 30 steps) of spatial HYP smearing to form a variational basis. Wilson loops were measured on 4646 gauge configurations, while Laplace trial states were measured on every fourth configuration only (1160 measurements). The error analysis in this work was done using the $\Gamma$ method \cite{Wolff:2003sm,Schaefer:2010hu} with a recent python implementation (pyerror)~\cite{Joswig:2022qfe} with automatic differentiation~\cite{Ramos:2020scv}.

\newpage

\begin{figure}[h]
\centering
\includegraphics[width=0.495\textwidth]{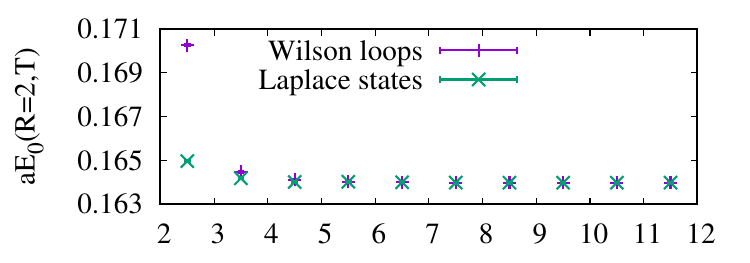}\\
\includegraphics[width=0.495\textwidth]{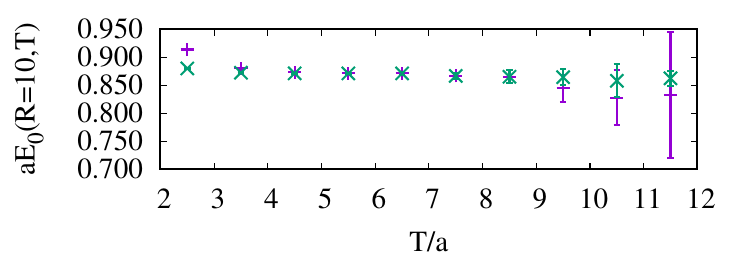}
\caption{The effective energies/masses using the Laplace trial states with  an optimal Gaussian profiles and Wilson loops with different HYP smearing levels for $R/a=2$ and $10$.}.
  \label{fig:exEm1}
\end{figure}

We compare the effective energies using the improved Laplacian eigenvector approach with Gaussian profiles after solving the GEVP together with smeared Wilson loop results in \fig{fig:exEm1}. Results from Laplacian modes show better ground state overlaps and higher accuracy than those from Wilson loops with only a 
quarter of the statistics. 

\begin{figure}[h]
\centering
\includegraphics[width=0.45\textwidth]{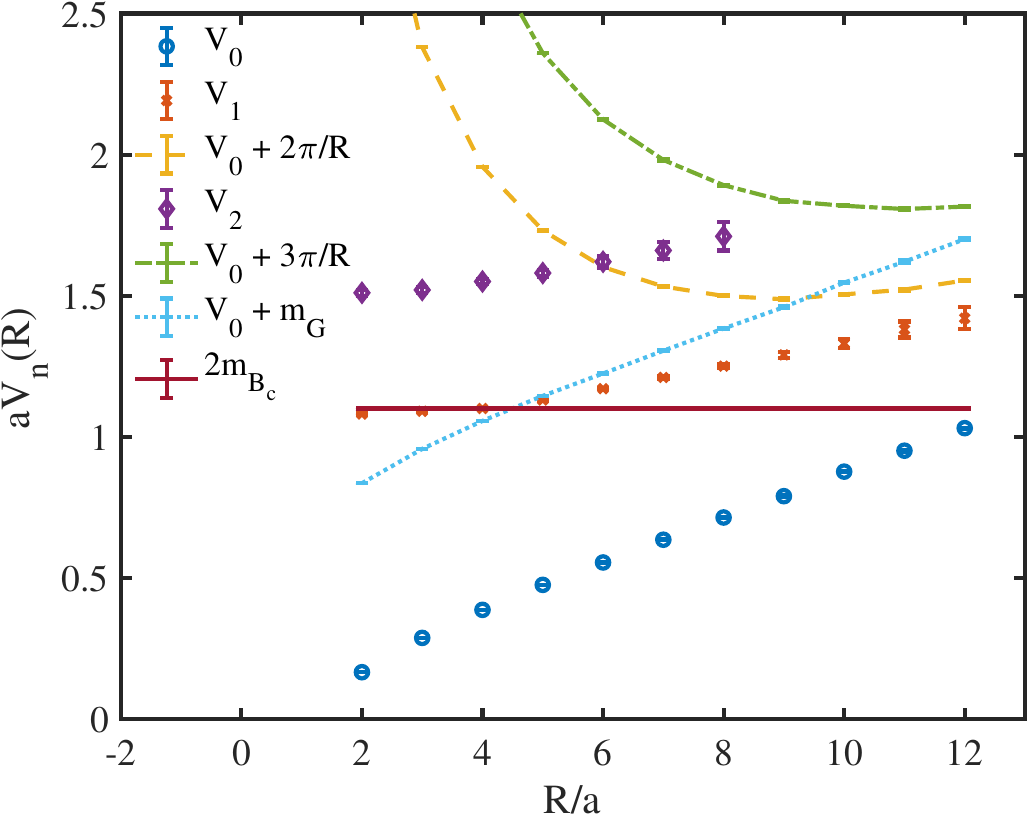}
\caption{The static potentials $V_n$ for the ground ($n=0$) and excited ($n=1,2$) states. We compare with radially excited string states $V_0+(n+1)\pi/R$, the lowest $0^{++}$ isoscalar meson (possible glueball) $V_0+m_G$ from~\cite{Urrea-Nino:2022gne} and two times the static-charm meson mass $2m_{B_c}$, also evaluated from Laplace trial states.}
  \label{fig:potEm1}
\end{figure}


In \fig{fig:potEm1} we present the static potentials $V_n$ for the ground ($n=0$) and excited ($n=1,2$) states using the Laplace trial states with optimal quark profiles after solving the GEVP. The excited states are just included to show the potential of the method, we want to stress here, that we only have the Laplace trial states in the operator basis, which just like Wilson loops may not have a good overlap with multi-particle states. Note, that we only analyze the $\Sigma_g^+$ state according to the nomenclature in \cite{Juge:1997nc, Juge:2002br} and its radial excitations, not the first-excited (hybrid) potential $\Pi_u$, lying between $\Sigma_g^+$ ($V_0$) and $\Sigma_g^+$' ($V_1$), which will be investigated in a future work, using covariant derivatives of eigenvectors in the trial states. For comparison we plot the radially excited string states $V_0+(n+1)\pi/R$, as well as the lowest $0^{++}$ isoscalar meson (possible glueball) $V_0+m_G$ from~\cite{Urrea-Nino:2022gne} and two times the static-charm meson mass $2m_{B_c}$. The latter was also evaluated using the new method, by combining our static perambulators $\tau_{ij}(\vec x,t_0,t_1)$ with a projector $P_+=(1+\gamma_0)/2$ and charm-quark perambulators $\tau_{ji}^{\alpha\beta}(t_1,t_0)=v_j^\dagger(t_1)[D^{-1}]_{t_1t_0}^{\alpha\beta}v_i(t_0)$ from~\cite{Knechtli:2022bji}, where the quark propagator $D^{-1}$ includes the dependence on the mass of the quark.

\begin{figure}[ht]
\centering
\includegraphics[width=0.495\textwidth]{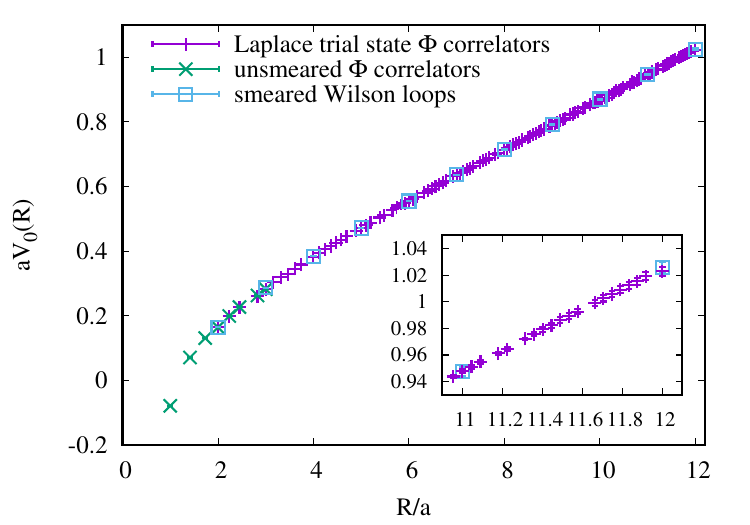}
\caption{The static ground state potential from optimal Lap- lace trial state correlators, computed for all on- and off-axis separations $R$ compared to on-axis Wilson loops. The green points for $R/a\leq3$ result from un-smeared Laplace trial state correlators (no HYP), showing the Coulomb behavior of the potential at small $R$, these are shifted vertically to match the potential with HYP2 smeared temporal links at $R/a=2$.}
  \label{fig:vpotEm1}
\end{figure}

The computational effort of this new method is less than the standard Wilson loop calculation, especially for off-axis separations. In fact, for our test ensemble on a $24^3\times48$ lattice the computation of on-axis Wilson loops using 4 spatial smearing levels (0, 10, 20, 30 HYP steps) is equally expensive as the calculation of 100 Laplacian eigenvectors and Laplace trial states with 7 Gaussian profiles including off-axis distances. The computational advantage of new method can be explained by the fact that the static perambulators can be computed first at each position, resulting in complex numbers, which then can easily be multiplied for arbitrary on- and off-axis separations without the need to compute spatial Wilson lines. 
In \fig{fig:vpotEm1} we present the optimal static potential $V_0(R)$ for all on- and off-axis separations $R/a$ from $N_v=100$ Laplacian eigenvectors compared to on-axis Wilson loop results, which agree well within errors. We also include a measurement of un-smeared Laplace trial state correlators for $R/a\leq3$ (no HYP smearing), showing the Coulomb behavior of the potential at small $R$. The green points in the plot are shifted vertically such that the un-smeared potential matches the potential with HYP2 smeared temporal links at $R/a=2$, which corresponds to removing the free energy difference. Further, we want to note that contrary to Wilson loops, Laplace trial states have an exact symmetry of the potential around half the lattice extension (in a specific direction $\vec r$), where in fact the force between $Q\bar Q$ must vanish due to the periodic boundary conditions, {\it i.e.}, the static potential should be flat. 

\section{The spatial distribution of optimal Laplace trial states}\label{sec:flux}

If we do not evaluate the spatial sum in the third line of the Laplace trial state in \eq{eq:lts}, we are left with an eigenvector  pair $v^\dagger(\vec z)v(\vec z)$ which acts as a 'test-charge' in the original Laplace trial state
\begin{eqnarray}
\hspace{-8mm}\psi^{(n)}(\vec z,R)=\hfill\\&&\hspace{-18mm}\bigg\langle\big|\big|\sum_{ij}^{N_v}\tilde\rho_R^{(n)}(\lambda_i,\lambda_j)v_i(\vec x)v_i^\dagger(\vec z)v_j(\vec z)v_j^\dagger(\vec x+R)\big|\big|_2\bigg\rangle\,,\non
\end{eqnarray}
which allows the scanning of individual contributions of the quark-anti-quark operator in a 3D time-slice via the free coordinate $\vec z$. We average over the whole lattice ($\vec x,t$), which already gives a very smooth signal on a single configuration. Note that we include the optimal trial state profiles  
\begin{eqnarray}
\tilde\rho_{R}^{(n)}(\lambda_i,\lambda_j)=\sum_{k,l}\nu_k^{(n)}u_{k,l}e^{-\lambda_i^2/4\sigma_l^2}e^{-\lambda_j^2/4\sigma_l^2}\,,
\end{eqnarray}
which in this case still depend on the two eigenvalues $\lambda_i$ and $\lambda_j$, since we did not perform the sum over $\vec z$ in \eq{eq:lts} and therefore did not get a $\delta_{ij}$. The singular vectors $u_k$ and generalized eigenvectors $\nu^{(n)}$ come from the SVD and GEVP in the static potential calculations for specific quark separation distances $R$ and allow us to look at the flux tube profiles for various energy states of $V_n(R)$. 

\begin{figure*}[ht]
$n=0:$\hfill.\\
\includegraphics[width=0.453\linewidth]{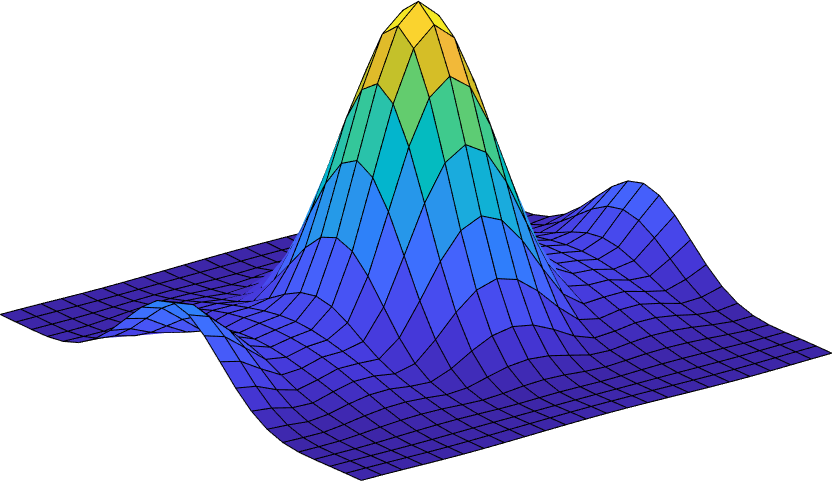}
\includegraphics[width=0.262\linewidth]{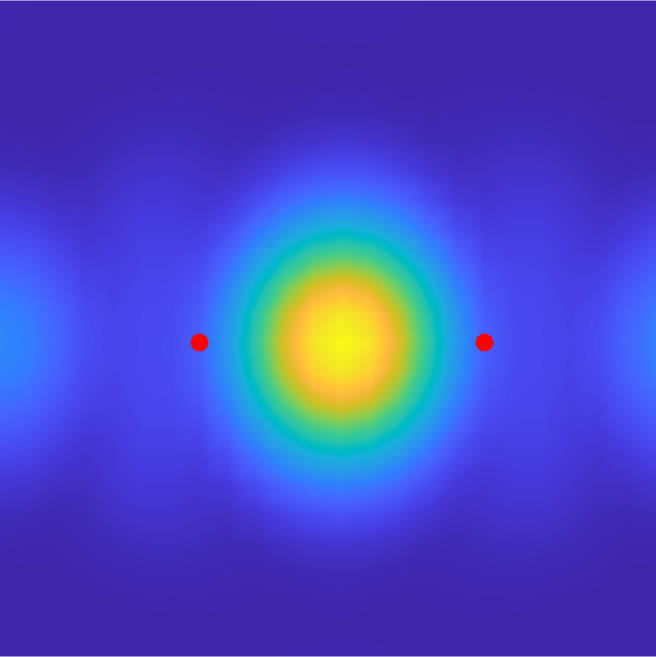}
\includegraphics[width=0.262\linewidth]{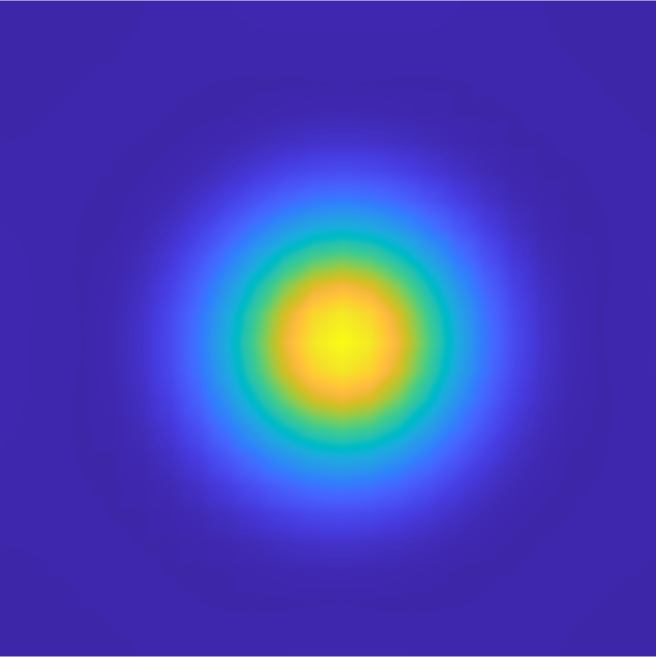}\\
$n=1:$\hfill.\\
\includegraphics[width=0.46\linewidth]{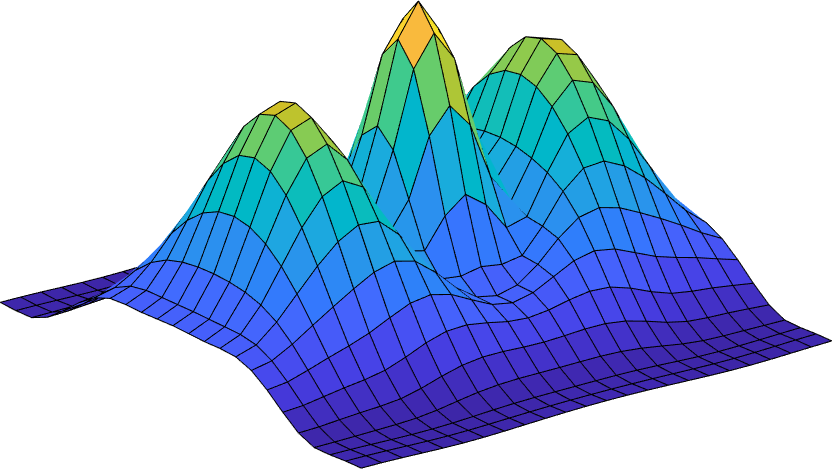}
\includegraphics[width=0.26\linewidth]{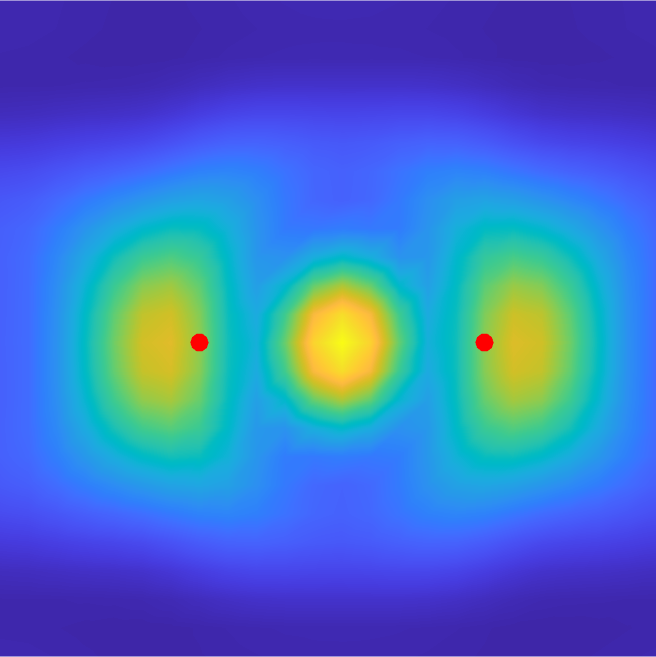}
\includegraphics[width=0.26\linewidth]{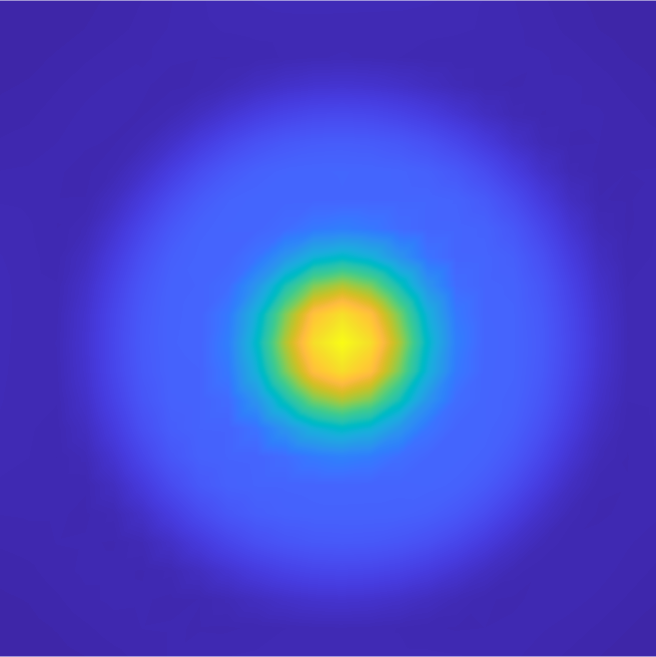}
\caption{Spatial distribution along and perpendicular to (right) the quark separation ($R=10a$) axis of the optimal Laplace trial state to measure the ground (top) and first excited (bottom) state potential of a static quark-anti-quark  pair, indicated by red dots.}
	\label{fig:r10ns}
\end{figure*}

In \fig{fig:r10ns} we present the spatial distributions of the optimal Laplace trial states to measure the ground resp. first excited state potentials of a static quark-anti-quark pair at spatial distance $R=10a$. The first excitation shows additional nodes in the spatial distribution along and perpendicular to the quark separation axis. The physical interpretation of these distributions in terms of the chromo-electromagnetic flux tube is not clear yet, the optimal profiles certainly contain some information of the ground and excited states of the static potential, the 'test-charge' $v(\vec z)v^\dagger(\vec z)$ however does not measure a specific color field component.

\section{Conclusions \& Outlook}\label{sec:co}

Alternative creation operators for static-quark-anti-quark states based on Laplacian eigenmodes are investigated. The use of a large number of eigenvectors weighted with Gaussian profiles is found to improve performance. An operator basis can be defined via different Gaussian profiles which can be analyzed with the GEVP formalism to extract optimal profiles and Laplace trial states. Temporal correlations of the new operators are used to compute static quark-anti-quark ground and excited state potentials. We observe earlier plateaus in the effective masses compared to standard Wilson loops. One significant advantage of the approach is its efficiency for computing the static potential not only for on-axis, but also for many off-axis quark-anti-quark separations. Indeed the new method requires far less computing time in particular for the latter case, since the eigenvector components of the covariant lattice Laplace operator have to be computed only once and can then be used for arbitrary on-axis and off-axis separations without the need to compute stair-like gauge-link connections. Finally, we visualize the spatial distribution of the optimal Laplace trial states for ground and excited state creation operators of the quark-anti-quark pair. 
We are currently working on an adaptation of the method to compute hybrid static potentials of exotic mesons, where gluonic string excitations requiring gluonic handles in the standard Wilson loop approach can be realized with covariant derivatives acting on the Laplacian eigenvectors, and to static-light mesons, cf. \cite{Bulava:2019iut}. First results were presented at the ConfinementXV~\cite{Hollwieser:2022bqp}, Lattice 2022~\cite{Hollwieser:2022pov} and ExcitedQCD~\cite{Hollwieser:2022app} conferences. 

\section*{Acknowledgements} The authors gratefully acknowledge the Gauss Centre for Supercomputing e.V. (www.gauss-centre.eu) for funding this project by providing computing time on the GCS Supercomputer SuperMUC-NG at Leibniz Supercomputing Centre (www.lrz.de). M.P. was supported by the European Union’s Horizon 2020 research and innovation programme under grant agreement 824093 (STRONG-2020). The work is supported by the German Research Foundation (DFG) research unit FOR5269 "Future methods for studying confined gluons in QCD". The project "Constructing static quark-anti-quark creation operators from Laplacian eigenmodes" is receiving funding from the programme " Netzwerke 2021", an initiative of the Ministry of Culture and Science of the State of Northrhine Westphalia, in the NRW-FAIR network, funding code NW21-024-A. The sole responsibility for the content of this publication lies with the authors. For valuable discussions we thank Pedro Bicudo and Jeff Greensite.


\bibliographystyle{utphys} 
\bibliography{paper}

\end{document}